\documentclass{article}
\usepackage{setspace}
\title{STRING DYNAMICS NEAR A KALUZA-KLEIN \\ BLACK HOLE}
\author{H.~K.~Jassal\thanks{E--mail : hkj@ducos.ernet.in},
A.~Mukherjee\thanks{E--mail : am@ducos.ernet.in} and
R.~P.~Saxena \thanks{Deceased}\\
	{\em Department of Physics and Astrophysics,} \\
	{\em University of Delhi, Delhi-110 007, India.} 
	}

\begin{document}
\doublespacing
\maketitle
\large
\newcommand{\del}{\mbox{$\partial$}} 
\begin{center}
\Large Abstract
\end{center}
The dynamics of a string near a Kaluza-Klein black hole are studied.
Solutions to the classical string equations of motion are obtained
using the world sheet velocity of light as an expansion parameter.
The electrically and magnetically charged cases are considered
separately.
Solutions for string coordinates are obtained in terms of the world-sheet
coordinate $\tau$.
It is shown that the Kaluza-Klein radius increases/decreases with
$\tau$ for electrically/magnetically charged black hole. \\

PACS number(s) : 04.50.+h, 11.25.Db, 04.70.Bw


\pagebreak
String propagation near a black hole is of great interest because  of
the interplay between the extended probe and the nontrivial
background geometry.
An extensive literature (for a review see \cite{erice}) deals with the
study of classical string dynamics in curved backgrounds.      
This investigation is important with a view to eventually understand
and interpret string quantization in curved spacetimes.    

A consistent quantization of strings requires string theory to be a
higher dimensional theory.   
The extra dimensions are compactified to obtain four-dimensional
spacetime. 
\footnote{Although string theory can be formulated directly in four dimensions
\cite{antoniadis}, the more popular approach is compactification of
the extra dimensions.}
It is an important question in string theory to study the mechanism of
this compactification.  
The extra dimensions are expected to contribute nontrivially
to the dynamics in the vicinity of a black hole, i.e., in the strong
gravity regime.  
An interesting approach could be to study how the compact extra
dimensions unfold as a string falls into a black hole.  
It is not unreasonable to hope that a string can be used as a probe
to understand how four-dimensional spacetime arises dynamically from
an underlying higher dimensional theory. 

The problem is complicated as it involves solving equations of motion
in $D$-dimensional ($D=26$ for bosonic strings and $D=10$ for
superstrings) spacetime, which includes the compact manifold.
As an {\it in-between} approach, we study string propagation in five-
dimensional Kaluza-Klein black hole backgrounds as a minimal extension
to four-dimensional curved spacetime.
These backgrounds are  solutions to the five-dimensional Einstein
equations, and include regular four-dimensional black hole solutions.

The string world-sheet action \cite{green} is given by 
\begin{equation}
S = -T_{0}\int d\tau d \sigma \sqrt{-detg_{ab}}
\end{equation}
where $g_{ab}=G_{\mu \nu}(X)\del_{a}X^{\mu}\del_{b}X^{\nu}$ is
the two-dimensional world-sheet metric; $\sigma$ and $\tau$ are the
world sheet coordinates.

The classical equations of motion in the
gauge $g_{ab}=\rho(\sigma,\tau)\eta_{ab}$ ($\eta_{ab}$ is the two-
dimensional Minkowskian metric) are given by 
\begin{equation}
\del_{\tau}^{2}X^{\mu}-c^{2}\del_{\sigma}^{2}X^{\mu}+\Gamma_{\nu
\rho}^{\mu} \left[\del_{\tau}X^{\nu}\del_{\tau}X^{\rho} -
c^{2}\del_{\sigma}X^{\nu}\del_{\sigma}X^{\rho}\right]=0 .
\end{equation}
and the constraints are given by
\begin{equation}
\del_{\tau}X^{\mu}\del_{\sigma}X^{\nu}G_{\mu \nu} = 0
\end{equation} 
\begin{equation}
[\del_{\tau}X^{\mu}\del_{\tau}X^{\nu} + c^{2}\del_{\sigma}X^{\mu}
\del_{\sigma}X^{\nu}]G_{\mu \nu} = 0. 
\label{cons2}
\end{equation}
\noindent Here $c$ is the velocity of wave propagation along the string.

Various simplifying ansatze exist for obtaining solutions to the
highly nonlinear system of equations (2)-(4), one being perturbation
expansion of string coordinates.    
We follow the approach of de Vega and Nicolaidis \cite{vega} which uses
the  world-sheet  velocity of light  as an expansion parameter. 
The scheme involves systematic expansion in powers of $c$.
If $c<<1$, the coordinate expansion is suitable to describe strings in
a strong gravitational background (see \cite{vega2, sanchez}).
Here, the derivatives w.r.t. $\tau$ overwhelm the $\sigma$ derivatives.  
In the opposite case ($c>>1$), the classical equations of motion give
us a stationary picture as the  $\sigma$ derivatives dominate. 

We restrict ourselves to the case where $c$ is small, our interest
being to probe the dynamical behaviour of the extra dimensions.  
The string coordinates are expressed as
\begin{equation}
X^{\mu}(\sigma, \tau) = A^{\mu}(\sigma, \tau) + c^{2} B^{\mu}(\sigma,
\tau) + c^{4} C^{\mu}(\sigma, \tau) + ..., 
\end{equation}
and the zeroth order $A^{\mu}(\sigma,\tau)$ satisfies the following
set of equations(with dot and prime denoting differentiation
w.r.t. $\tau$ and $\sigma$ respectively);
\begin{eqnarray}
\ddot A^{\mu} + \Gamma_{\nu \rho}^{\mu} \dot A^{\nu} \dot A^{\rho} & =
& 0, \\ \nonumber
\dot A^{\mu} \dot A^{\nu} G_{\mu \nu} & = & 0, \\ \nonumber
\dot A^{\mu} A'^{\nu} G_{\mu \nu} & = & 0.
\end{eqnarray}
These equations describe the motion of a null string \cite{vega}. 
The second constraint is the 'stringy' constraint and restricts the
motion to be perpendicular to the string.

The metric background to study string propagation
\cite{gibbons} (see also \cite{miriam}) is   
\begin{equation}
ds^{2}=-e^{4 k \frac{\varphi}{\sqrt{3}}} (dx_{5} +
2 k A_{\alpha}dx^{\alpha})^{2} + e^{-2 k \frac{\varphi}{\sqrt{3}}}g_{\alpha
\beta}dx^{\alpha}dx^{\beta},
\end{equation}
where $k^{2}=4 \pi G$; $x_{5}$ is the extra dimension and should be
identified modulo $2\pi R_0$, where $R_0$ is the radius of the circle
about which the coordinate $x_5$ winds.
The quantity $R_0$ is the asymptotic value of a dynamical quantity,
the Kaluza-Klein radius, which is discussed below.
Here $g_{\alpha \beta}$ is the four-dimensional spacetime.

The mass $M$ of the black hole, the electric charge $Q$ and the
magnetic charge $P$ are related to the scalar charge $\Sigma$ by  
\begin{equation}
\frac{2}{3} \Sigma = \frac{Q^{2}}{\Sigma + \sqrt{3} M} +
\frac{P^{2}}{\Sigma - \sqrt{3} M},
\end{equation}
where the scalar charge is defined as  \\
\begin{center}
$k\varphi \longrightarrow \frac{\Sigma}{r} +
O\left(\frac{1}{r^2}\right)$ as $r \longrightarrow \infty.$ 
\end{center}

We follow the notation of ref. \cite{itzhaki}.
The black hole solutions are
\begin{eqnarray}
e^{4 \varphi/\sqrt{3}} & = & \frac{B}{A}, A_{\mu}dx^{\mu} =
\frac{Q}{B}(r-\Sigma)dt + P \cos\theta d\phi \\ \nonumber
g_{\mu \nu}dx^{\mu}dx^{\nu} & = & \frac{f^2}{\sqrt{AB}} dt^{2} -
\frac{\sqrt{AB}}{f^2} dr^2 - \sqrt{AB} \left(d\theta^2 + \sin^2\theta
d\phi^2\right) 
\end{eqnarray}
Here $A$, $B$ and $f$ are given by

\begin{eqnarray}
A & = & (r-\frac{\Sigma}{\sqrt{3}})^{2} - \frac{2
P^{2}\Sigma}{\Sigma-\sqrt{3} M}  \\ \nonumber
B & = & (r+\frac{\Sigma}{\sqrt{3}})^{2} - \frac{2 Q^{2} \Sigma}{\Sigma
+ \sqrt{3} M}\\ \nonumber
f^{2} & = & (r-M)^{2} - (M^{2} + \Sigma^{2} - P^{2} - Q^{2}) 
\end{eqnarray}

We seek to solve the equations of motion for the string coordinates in
the exterior of the black hole. 
For simplicity, we consider the magnetically and electrically charged
cases separately. 
The zeroth order equations of motion for string coordinates in the
electrically neutral ($Q=0$) background are 
\begin{eqnarray}
\frac{\del^{2}t}{\del \tau^{2}} + 2\left(\frac{f'}{f} -
\frac{B'}{2B}\right) \frac{\del t}{\del \tau} \frac{\del
r}{\del \tau} &=&0, \\ \nonumber
\frac{\del^{2}r}{\del \tau^{2}} + \left[
-\frac{f^{3}}{2AB^{2}}(B'f-2f'B) \right] \left( \frac{\del
t}{\del \tau} \right)^{2} + 
\left(\frac{A'f -  2f'A}{2Af} \right)
\left( \frac{\del r}{\del \tau} \right)^{2}&-& \\ \nonumber 
\frac{f^{2}A'}{2A} \left( \frac{\del \phi}{\del \tau}
\right)^{2} + 
\frac{f^2}{2A^{3}}(A'B-B'A) \left( \frac{\del
x_{5}}{\del \tau} \right)^{2} &=&0, \\ \nonumber
\frac{\del^{2}\phi}{\del \tau^{2}} + \frac{A'}{A}
\left(\frac{\del r}{\del \tau} \right) \left(\frac{\del
\phi}{\del \tau} \right) &=& 0, \\ \nonumber
\frac{\del^{2} x_{5}}{\del \tau^{2}} + \left(-\frac{A'}{A}
+ \frac{B'}{B} \right) \left(\frac{\del r}{\del
\tau}\right)\left(\frac{\del x_{5}}{\del \tau}\right) &=& 0.  
\end{eqnarray}
Here we restrict ourselves to the exterior region $r>M$ and the
equatorial plane, i.e. $\theta=\pi/2$. 

The first integrals of motion are given by
\begin{eqnarray}
\frac{\del t}{\del \tau} &=&  \frac{c_{1}B}{f^{2}}, \\ \nonumber
\frac{\del \phi}{\del \tau} &=& \frac{c_{2}}{A},	 \\ \nonumber
\frac{\del x_{5}}{\del \tau} &=& c_{3}\frac{A}{B}		 \\ \nonumber
\left(\frac{\del r}{\del \tau}\right)^2 &=& \frac{B}{A}
c_{1}^2 - \frac{f^2}{A^2} c_{2}^2 - \frac{f^2}{B}c_{3}^2.
\end{eqnarray}

where $c_{1}$, $c_{2}$ and $c_{3}$ are integration constants.

The equations can be further reduced to quadratures
\begin{eqnarray}
\tau &=& \int \frac{dr}{\sqrt{\frac{B}{A}c_{1}^2 - \frac{f^2}{B}
c_{3}^2}}, \\ \nonumber 
x_{5} &=& \int \frac{c_{3}A dr}{B\sqrt{\frac{B}{A}c_{1}^2 -
\frac{f^2}{B} c_{3}^2}}, \\ \nonumber 
t &=& \int \frac{ c_{1} B dr}{f^2 \sqrt{\frac{B}{A}c_{1}^2 -
\frac{f^2}{B} c_{3}^2}}, 
\label{integs}
\end{eqnarray}
up to constants of integration which depend on $\sigma$ and can be
solved numerically to obtain $t$, $r$, $\phi$ and $x_{5}$ as functions
of $\tau$. 
Here, we have taken $c_{2}=0$, i.e. the string is falling in 'head-on'
and we work in the region where $r>>\Sigma$.

The integrals (\ref{integs}) have been evaluated numerically and inverted
to obtain the coordinates as function of $\tau$.
The behaviour of $r$ as a function of $\tau$ is shown in Fig. 1 and
that of the coordinate $x_5$ is shown in Fig. 2 for different values
of $c_1$ and $c_3$.

For the electrically charged ($P=0$) black hole, the equations of
motion in the zeroth order take the form  
\begin{eqnarray}
AB[Af^2+12 Q^2(r-\Sigma_{1})^2]\frac{\del^{2}t}{\del \tau^2}
-[12A'BQ^2(r-\Sigma_{1})^2+B'A^2f^2 &-&  \\ \nonumber 
4B'AQ^2(r-\Sigma_{1})^2-2f'A^2Bf-8AB Q^2 
(r-\Sigma_{1})]\frac{\del t}{\del \tau} \frac{\del r}{\del \tau} &+&
\\ \nonumber
4QAB[B'(r-\Sigma_1)-B] \frac{\del r}{\del \tau} \frac{\del x_5}{\del
\tau} &=& 0,   \\ \nonumber
2A^3B^2f\frac{\del^2 r}{\del
\tau^2} + f^3[4A'BQ^2(r-\Sigma_1)^2-B'A^2f^2 
+ 4B'AQ^2(r-\Sigma_1)^2 &+&  \\ \nonumber2f'A^2Bf-8ABQ^2(r-\Sigma_1)]
\left(\frac{\del t}{\del \tau} \right)^2 -
8f^3B^2Q[A'(r-\Sigma_1)-A]
\frac{\del t}{\del \tau} \frac{\del x_5}{\del \tau} &+& \\ \nonumber 
 A^2B^2[A'f-2f'A] \left(\frac{\del r}{\del \tau}\right)^2
-A^2B^2f^3A'\left(\frac{\del \phi}{\del \tau}\right)^2 +
f^3B^2[A'B-B'A] \left(\frac{\del x_5}{\del \tau}\right)^2 &=& 0, \\ \nonumber
AB^2[Af^2 + 12Q^2(r - \Sigma_{1})^2]\frac{\del^{2}x_5}{\del
\tau^2}-4QA[A'Bf^2(r-\Sigma_1) -
B'Af^2(r-\Sigma_1) &+&  \\ \nonumber
4B'Q^2(r-\Sigma_1)^3+2f'ABf(r-\Sigma_1)-ABf^2 -
4BQ^2(r-\Sigma_1)^2] \frac{\del t}{\del \tau} \frac{\del r}{\del \tau}
&-& \\ \nonumber B[A'ABf^2+12A'BQ^2(r-\Sigma_1)^2-B'A^2f^2 +
4B'AQ^2(r-\Sigma_1)^2 &-& \\ \nonumber16ABQ^2(r-\Sigma_1)]
\frac{\del r}{\del \tau} \frac{\del x_5}{\del \tau} &=& 0.
\label{elec}
\end{eqnarray}
where $\Sigma_1=\Sigma/\sqrt{3}$ and  the constraint equation is
\begin{eqnarray}
\left\{ \frac{f^2}{B}-\frac{4Q^2}{AB}(r-\Sigma_1)^2 \right\}\left(
\frac{\del t}{\del \tau} \right)^2 - \frac{8Q}{A}(r-\Sigma_1)
\frac{\del t}{\del t} \frac{\del x_5}{\del \tau} &-& \\ \nonumber
- \frac{A}{f^2} \left( \frac{\del r}{\del \tau} \right)^2 -A \left(
\frac{\del \phi}{\del \tau} \right)^2 -\frac{B}{A} \left( \frac{\del
x_5}{\del \tau} \right)^2 &=& 0.
\label{constr}
\end{eqnarray}

The structure of the equations in this case is such that they are  not
reducible to quadratures and we have to resort to solving them
numerically.  
Again, we consider an infalling string in the region where $r>>\Sigma$
and $\theta=\pi/2$.  
 
The set of equations (\ref{elec}) and (\ref{constr}) has been solved
numerically to obtain the coordinates as functions of $\tau$.
Again we have a two parameter family of solutions.
Fig. 3 and Fig. 4 illustrate how $r$ and $x_5$ vary as functions of
$\tau$ for different choices of integration constants.

The coordinate $x_5$ increases monotonically in the magnetic case, while
in the electric case, it first increases and then starts decreasing. 
The two cases indicate, in the magnetic case, that the coordinate
$x_5$ goes about a circle continuously in one direction, while in the
electric case the direction reverses.
Although the behaviour is different in the two cases, the picture
becomes clear if we define a quantity, the Kaluza-Klein radius, which
is related to its asymptotic value $R_0$ as
\begin{equation}
R(r)  =  R_{0} \left( \frac{B}{A} \right)^{1/2}. 
\end{equation}
The radius $R(r)$ is a dynamical quantity as it depends implicitly on
$\tau$ through $R(\tau)=R(r(\tau))$. 
The effect of the magnetic field is to shrink the extra dimension (as
already mentioned in \cite{gibbons}) i.e., as the string approaches
the black hole, the value of the Kaluza-Klein radius which it sees
becomes smaller than its asymptotic value.  
The presence of electric charge tends to expand the extra dimension.
Fig. 5 and Fig. 6 clearly show that the behaviour of the Kaluza-Klein
radius is  opposite in the electrically and magnetically charged
cases. 

We have studied propagation of a null string in five-dimensional,
electrically and magnetically charged,  Kaluza-Klein black hole
backgrounds. 
Our study of string propagation in Kaluza-Klein backgrounds is
motivated by the importance of such backgrounds in the context of
toroidal compactification schemes for string theory.
Here, we have tried to explore the behaviour of the extra {\it fifth}
dimension as the string approaches the black hole horizon.
The solutions we have obtained are valid in the region outside the
horizon but not asymptotically far from the horizon.

The solutions, in the limit $\Sigma \longrightarrow 0$, match with the
ones given in \cite{sanchez}.
The essential difference lies in the presence of the extra dimension. 
Another paper that deals with five-dimensional Kaluza-Klein black
holes \cite{itzhaki} finds out string corrections to the five-dimensional
vacuum Einstein equations and their effect on the black hole metrics.
On the other hand our approach is to study the dynamics of a string
probe in a classical background.
The two approaches are complementary to each other.
 
In the present paper, we have only considered the classical picture. 
In principle, however, one expects quantum effects to be dominant in
the strong gravity regime. 
Nevertheless, we expect the classical picture to give an intuitive
idea of the mechanism of compactification. 
It is clear from the above considerations that, even in the classical
regime, we can probe the expansion/shrinking of the compact
dimension. 
However, the effects of the 
background (and hence the compact dimension) on the string probe
itself, in terms of changes in its shape and conformation, are
identically zero in the zeroth order of the expansion in $c$.
These effects are expected to manifest themselves if we go to higher
orders in $c$. 
Work on this is in progress and will be reported elsewhere.

H.~K.~J. thanks the University Grants Commission, India, for a
fellowship.

\pagebreak
\begin{figure}[ht]
\vskip 15truecm
\includegraphics{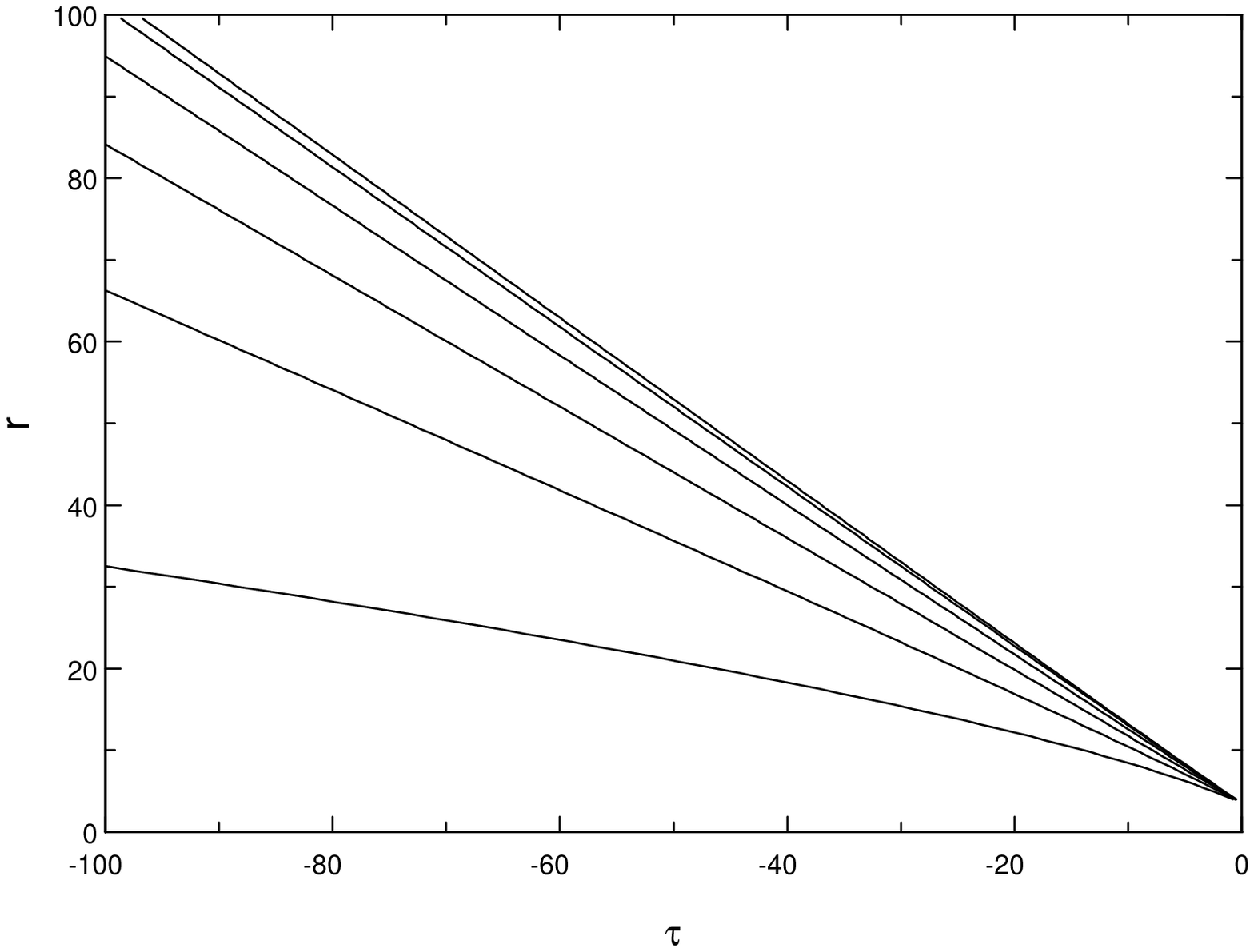}
\caption{Plot of $r$ vs.$\tau$ for magnetically charged black hole, for
different choices of $c_1$ and $c_3$..}
\end{figure}
\begin{figure}[ht]
\vskip 15truecm
\includegraphics{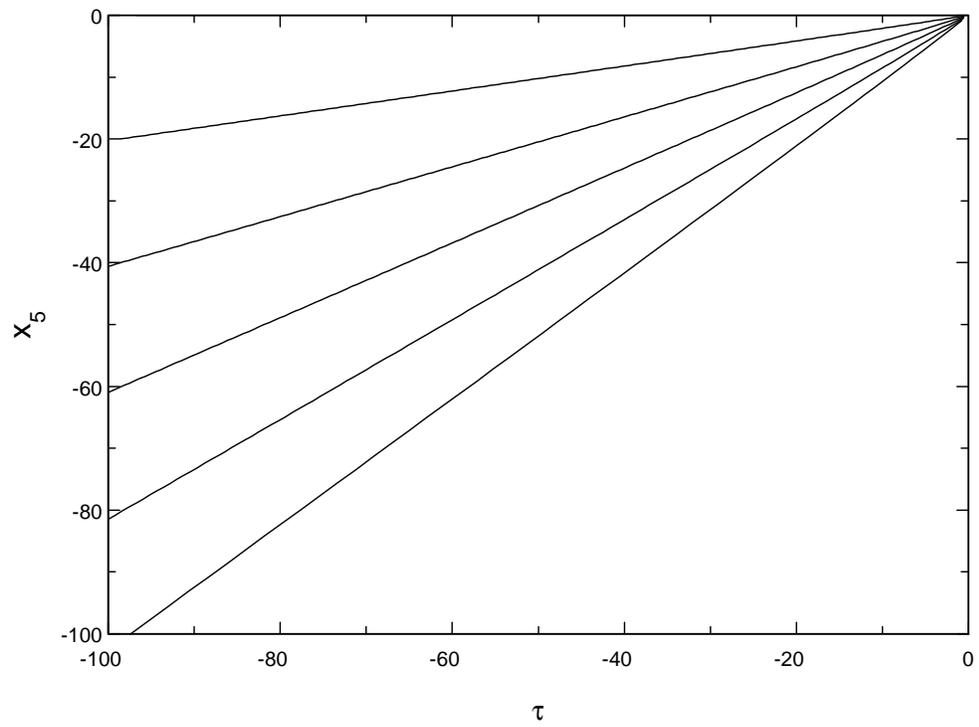}
\caption{$x_5$ vs. $\tau$ for magnetically charged black hole.}
\end{figure}
\begin{figure}[ht]
\vskip 15truecm
\includegraphics{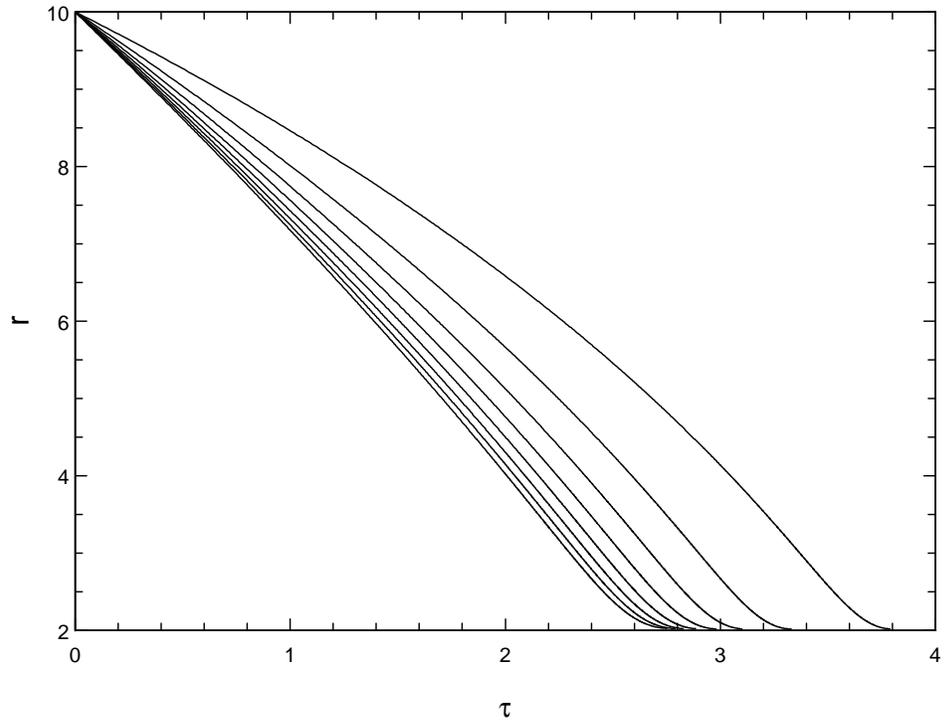}
\caption{Plot of $r$ w.r.t. $\tau$ for electrically charged black
hole, for different choices of integration constants. }
\end{figure}
\begin{figure}[ht]
\vskip 15truecm
\includegraphics{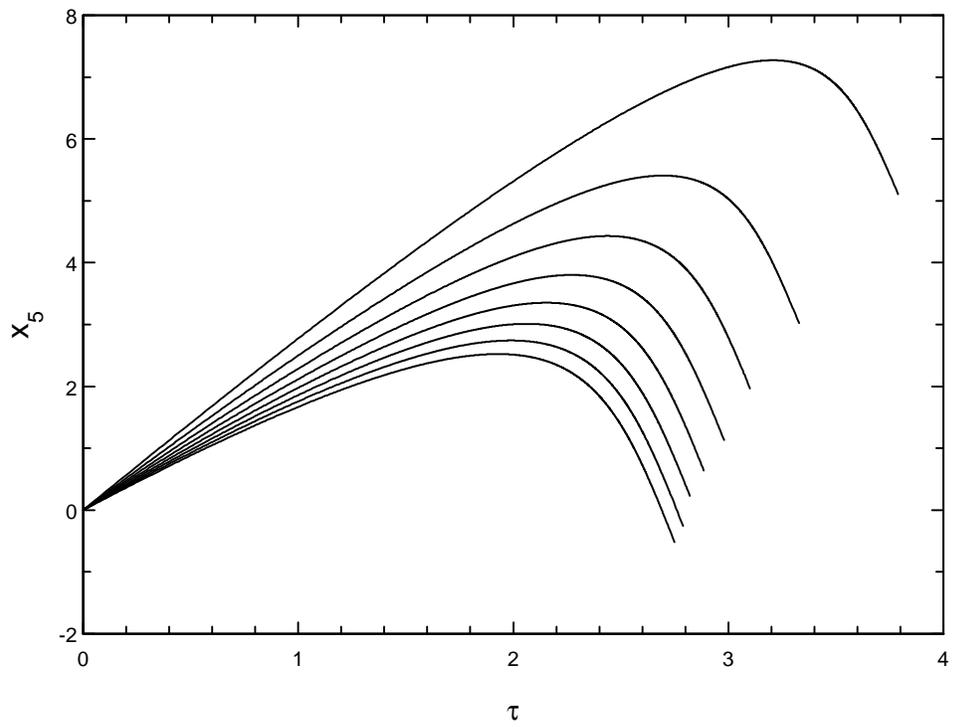}
\caption{$x_5$ vs. $\tau$ for electrically charged black hole.}
\end{figure}
\begin{figure}[ht]
\vskip 15truecm
\includegraphics{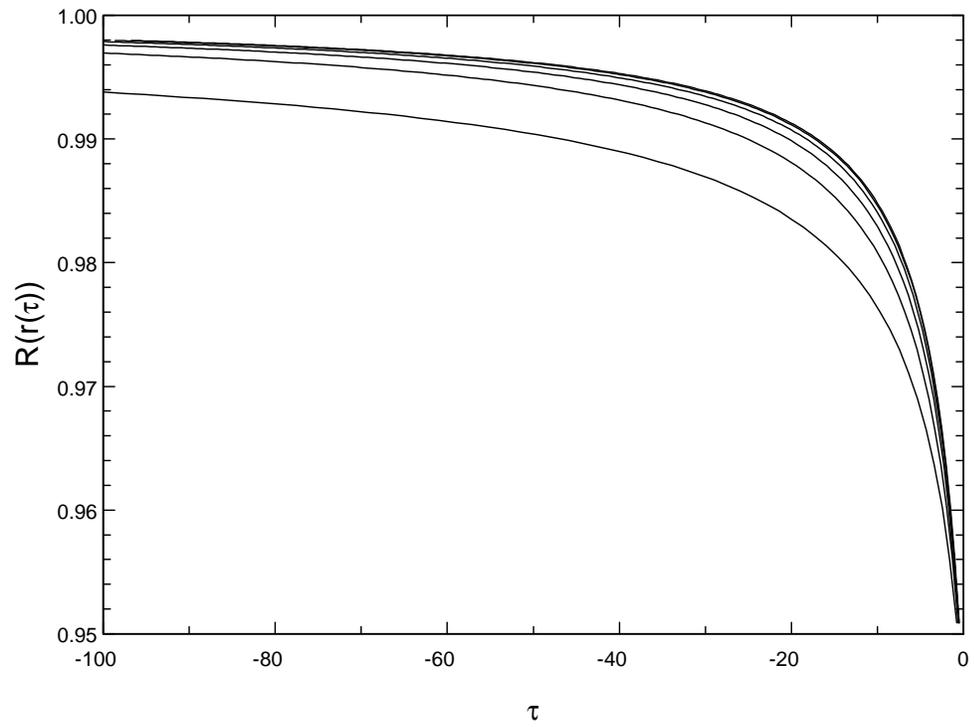}
\caption{Kaluza-Klein radius as a function of $\tau$ for magnetic black hole.}
\end{figure}
\begin{figure}[ht]
\vskip 15truecm
\includegraphics{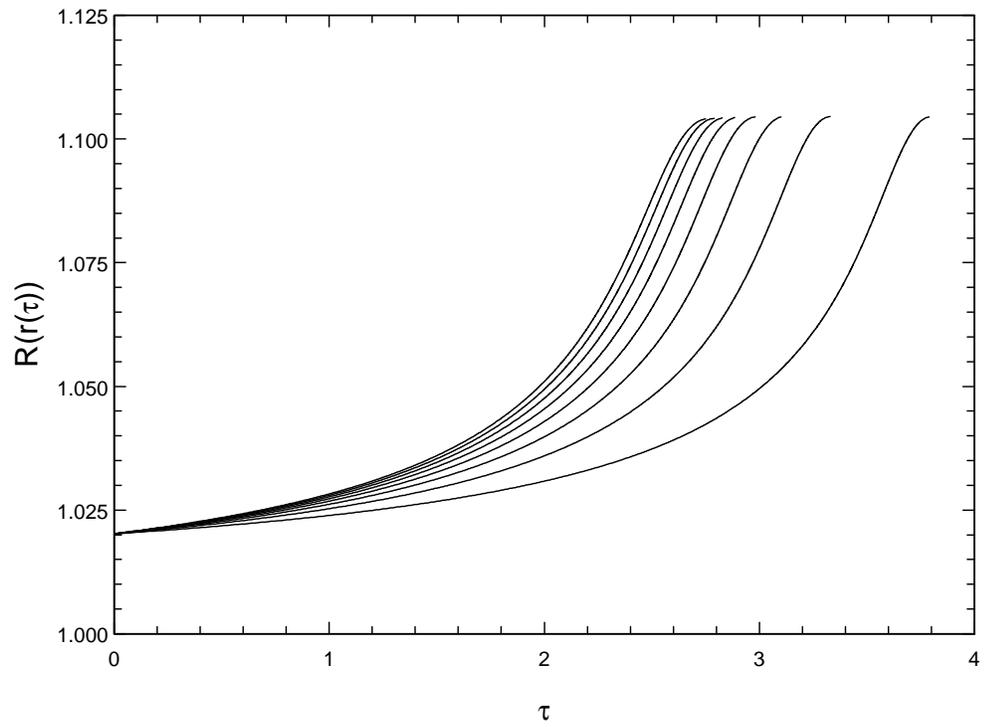}
\caption{Kaluza-Klein radius for electrically charged case.}
\end{figure}
\end{document}